\documentclass[letterpaper, 12 pt, conference]{ieeeconf}  

\onecolumn

\IEEEoverridecommandlockouts                              
\usepackage[utf8]{inputenc}
\usepackage[T1]{fontenc}

\usepackage{graphicx} 
\usepackage{color}
\usepackage{multicol}
\usepackage{multirow}
\usepackage[T1]{fontenc}
\usepackage{threeparttable}

\title{\LARGE \bf
A System-on-Chip for Closed-loop Optogenetic Sleep Modulation 
}

\author{Xilin Liu$^{1}$ and Andrew G. Richardson$^{2}$
\thanks{$^{1}$Xilin Liu is with the Department of Electrical and Computer Engineering, University of Toronto, ON M5S 3G4, Canada
        {\tt\small xilinliu at ece.utoronto.ca}}%
\thanks{$^{2}$Andrew G. Richardson is with the Department of Neurosurgery, University of Pennsylvania, PA 19104, USA
        {\tt\small Andrew.Richardson at pennmedicine.upenn.edu}}%
}

\begin{document}

\noindent Accepted for publication at the International Conference of the IEEE Engineering in Medicine and Biology Society (EMBC), 2021

{\let\newpage\relax\maketitle}

\thispagestyle{empty}
\pagestyle{empty}

\begin{abstract}
Stimulation of target neuronal populations using optogenetic techniques during specific sleep stages has begun to elucidate the mechanisms and effects of sleep. To conduct closed-loop optogenetic sleep studies in untethered animals, we designed a fully integrated, low-power system-on-chip (SoC) for real-time sleep stage classification and stage-specific optical stimulation. The SoC consists of a 4-channel analog front-end for recording polysomnography signals, a mixed-signal machine-learning (ML) core, and a 16-channel optical stimulation back-end. A novel ML algorithm and innovative circuit design techniques improved the online classification performance while minimizing power consumption. The SoC was designed and simulated in 180 nm CMOS technology. In an evaluation using an expert labeled sleep database with 20 subjects, the SoC achieves a high sensitivity of 0.806 and a specificity of 0.947 in discriminating 5 sleep stages. Overall power consumption in continuous operation is 97 $\mu$W.
\end{abstract}

\section{Introduction}

Sleep provides numerous physiological and cognitive benefits. Consequently, scientists endeavor to both understand the neural circuits that control sleep and develop interventions that target these circuits to treat sleep disorders and enhance benefits. Sleep is not a homogenous state; it is composed of a sequence of rapid eye movement (REM) sleep and three stages of non-REM sleep (N1, N2, and N3). Each stage has a signature of electrical activity in the brain (EEG), muscles (EMG), and eyes (EOG), collectively called the polysomnogram (PSG), which is used for its classification.

\begin{figure}[!ht]
  \centering
  \includegraphics[width=0.4\textwidth]{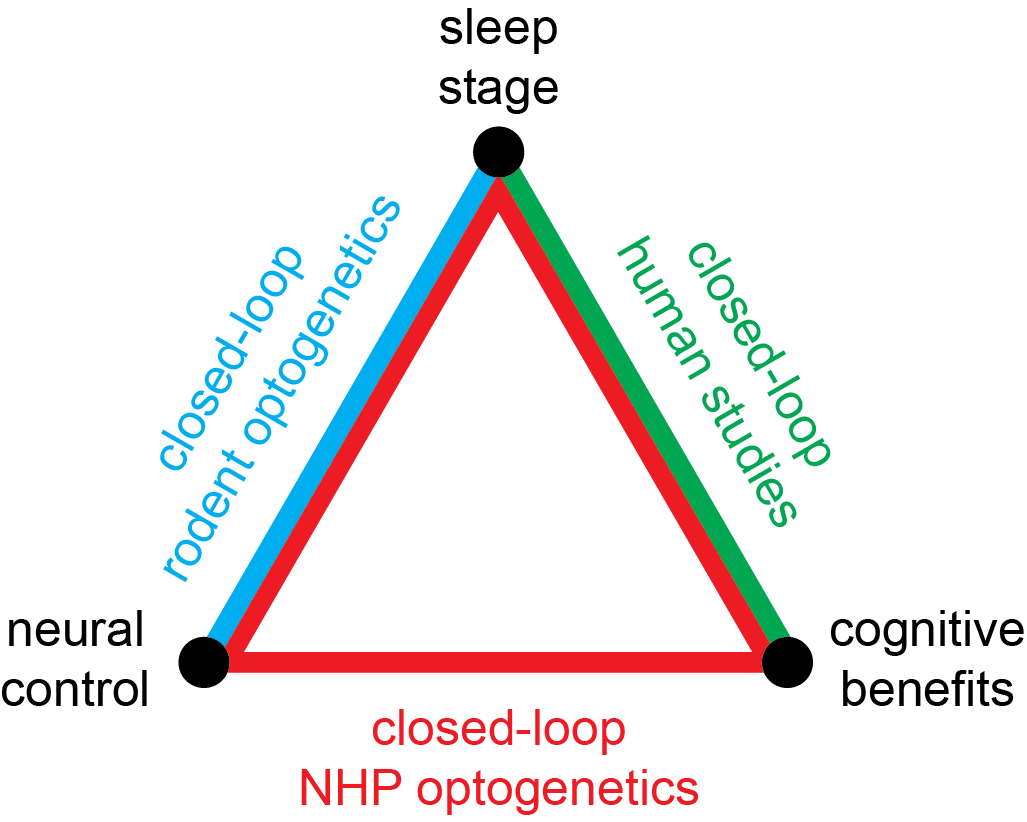}\\
  \caption{Summary of how closed-loop studies in different animal models (rodent, nonhuman primate, human) can assess causal relationships between sleep stages, controlling neural circuits, and resulting cognitive benefits.}\label{fig_intro}
\end{figure}

Sleep interventions and causal investigations increasingly rely on closed-loop paradigms in which a real-time sleep stage classifier is used to deliver stage-specific stimulation. For example, in humans, auditory or noninvasive electrical stimulation applied in phase with the prominent 1-Hz oscillatory brain activity of stage N3 can enhance long-term memory \cite{Ngo2013,Ketz2018}. However, closed-loop human studies linking sleep stages to cognitive benefits (Fig. \ref{fig_intro}, green) suffer from irreproducibility \cite{Henin2019,Bueno2019}, perhaps in part due to the rather nonspecific stimulation.

More precise interventions can be performed using optogenetics, where the activity of genetically-modified neurons is controlled by light \cite{Weber2016}. Closed-loop optogenetic studies in transgenic mice have identified circuits controlling sleep. For example, activation of orexin or noradrenergic neurons during REM or N3 evokes transition to wakefulness \cite{Carter2010}. However, while these closed-loop rodent studies link sleep stages to the underlying neuronal populations (Fig. \ref{fig_intro}, blue), limitations in the cognitive abilities of mice restrict assessment of the effects of sleep on cognition.

Unlike mice and humans, an animal model capable of higher-level cognitive function that is amenable to optogenetic techniques could directly test hypotheses linking cognitive effects to neural circuits controlling specific sleep stages (Fig. \ref{fig_intro}, red). Nonhuman primates (NHPs) meet these requirements \cite{Roelfsema2014,Tremblay2020}. However, due to their size, strength, and dexterity, freely behaving and sleeping NHPs cannot be tethered to a data acquisition, classification, and stimulation system for closed-loop experiments. This hardware needs to be embedded on the animal. Very little sleep research is conducted in NHPs due to this technological barrier.

Therefore, in this work, we developed a system-on-chip (SoC) for closed-loop optogenetic sleep research in untethered subjects. The SoC was designed and simulated in 180nm CMOS technology. In a simulation with 40 nights of recordings from 20 subjects, this design achieves a sensitivity of 0.806 and a specificity of 0.947 in discriminating the 5 sleep stages (wake, REM, N1, N2, N3).

\section{Design}

\begin{figure}[!ht]
  \centering
  \includegraphics[width=0.7\textwidth]{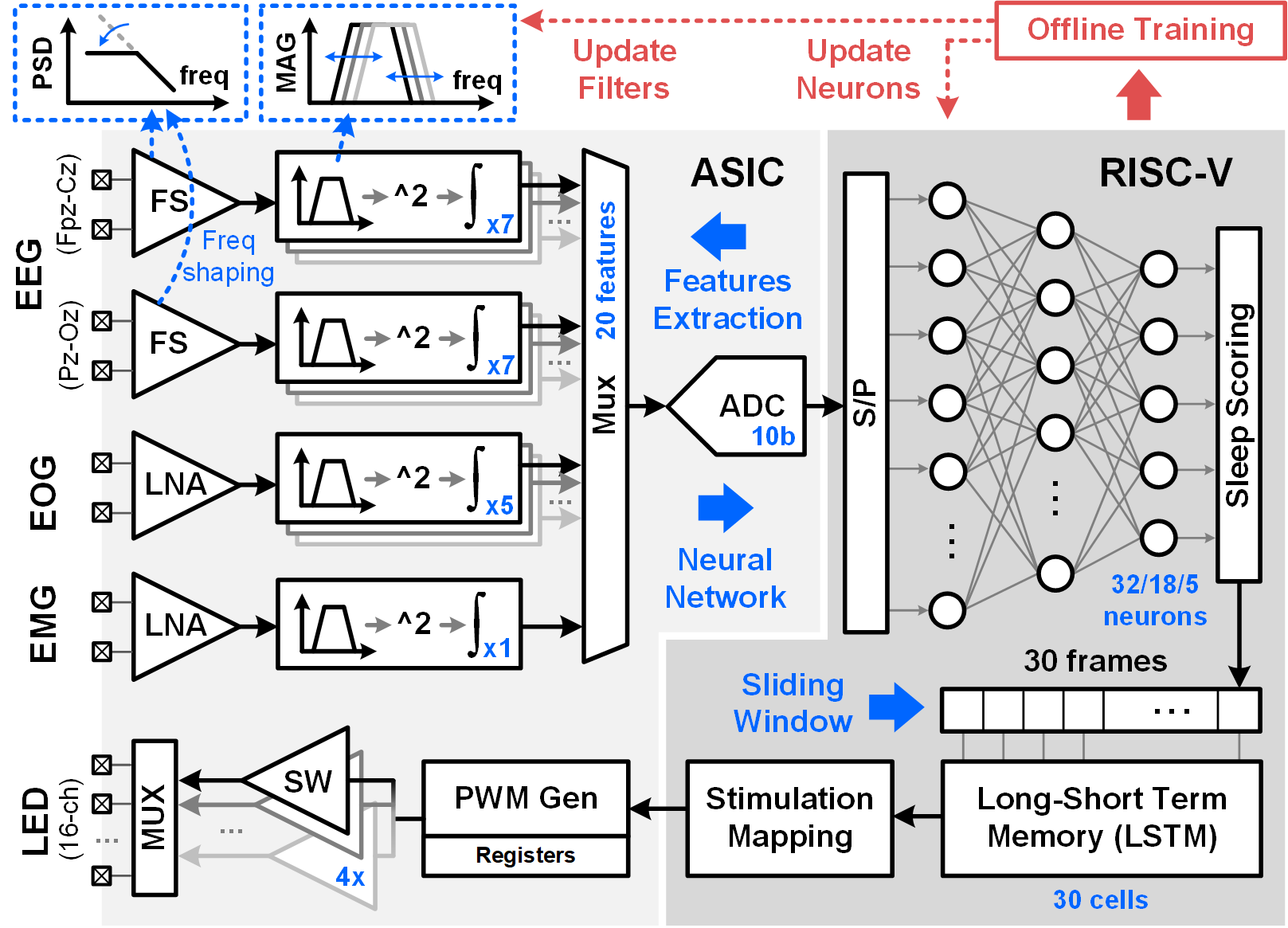}\\
  \caption{System diagram of the SoC for closed-loop optogenetic sleep research. The SoC consists of an ASIC for bidirectional neural interface and neural feature extraction and a RISC-V for NN-based sleep stage classification and closed-loop control.}\label{fig_sys_diagram}
\end{figure}

Fig. \ref{fig_sys_diagram} depicts the system diagram of the SoC. The SoC consists of an application-specific integrated circuit (ASIC) and an open-source RSIC-V core. The ASIC integrates 4 channels of neural amplifiers, 20 channels of feature extraction units, and 4 channels of optical stimulators. The overall closed-loop operating principle is as follows. The neural amplifiers acquire 4 PSG signals including 2 channels of EEG, 1 channel of EOG, and 1 channel of EMG. Narrow-band energy features are extracted from the amplified signals in the 20 feature extraction channels with programmable frequency bins spaced in the natural logarithmic domain. The time constant of the energy integral is 1 sec. The 20 energy features are digitalized in a 10-bit successive approximation register (SAR) analog-to-digital converter (ADC), and the results are sent serially to the two-stage neural network (NN) implemented in the RSIC-V. The NN consists of a first stage that scores the sleep stages from the extracted features and a second stage that classifies the sleep stages from a succession of 30 outputs from the first stage. The final detected sleep stage is used to trigger a pre-defined optical stimulus controlled by a pulse-width modulation (PWM) signal. The 4-channel stimulators are connected to an off-chip LED array via a multiplexer. The simulated overall system power is 97$\mu$W during a typical continuous operation.

\begin{figure}[!ht]
  \centering
  \includegraphics[width=0.5\textwidth]{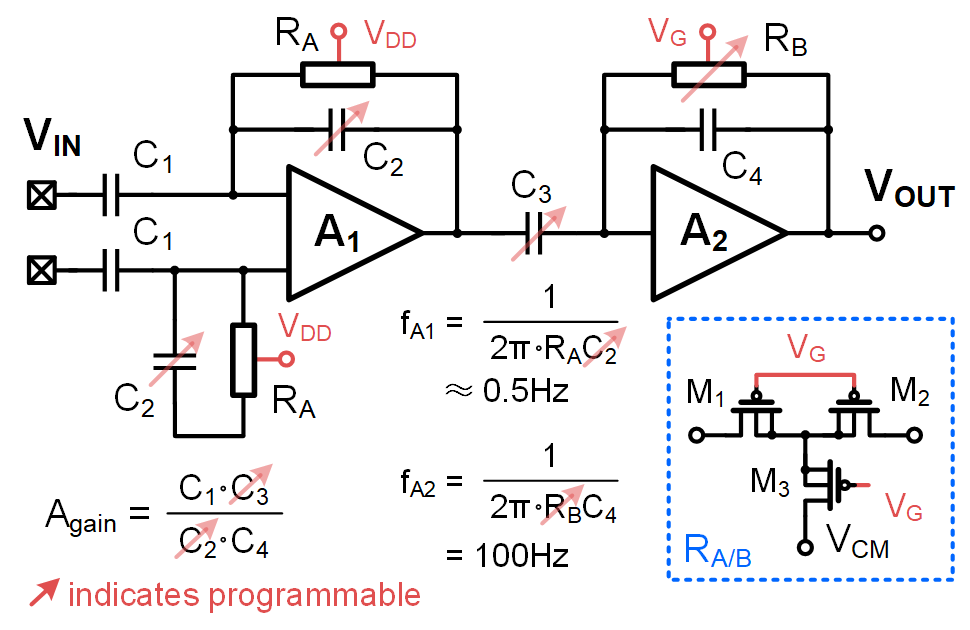}\\
  \caption{Schematic of the frequency shaping neural amplifier.}\label{ckt_neural_amplifier}
\end{figure}

A pre-whitening technique was used in the EEG amplifier. The power spectrum of EEG has a $(1/f)^n$ characteristic, where n is between 2 to 4, while the noise floor of a typical neural amplifier follows $1/f$ in the frequency range of interest \cite{Miller2011}. Thus, the signal-to-noise ratio (SNR) increases as frequency decreases, which indicates that an amplifier can be designed with a spectral whitening without scarifying the overall SNR. Moreover, such whitening normalizes the extracted energy features of different frequency bands, which is beneficial for the NN. Simple pre-whitening can be implemented by adding one dominant pole in the transfer function, as illustrated in Fig. \ref{ckt_neural_amplifier}. In this work, a high-pass frequency of 100Hz was used. The core low-noise amplifier ($A_1$) uses a complementary input MOSFET differential pair biased in the subthreshold region for maximizing power and noise efficiency.

\begin{figure}[!ht]
  \centering
  \includegraphics[width=0.6\textwidth]{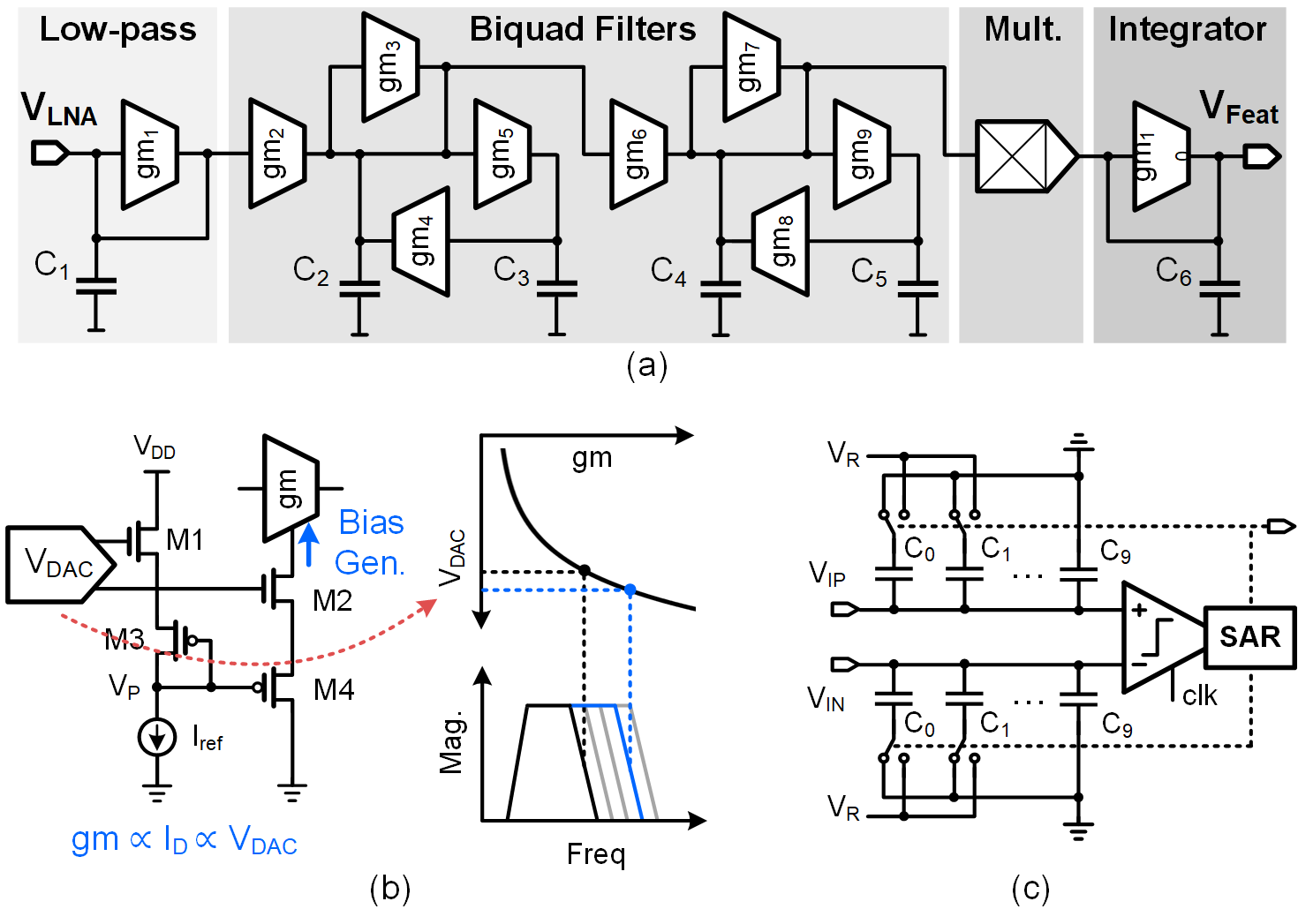}\\
  \caption{(a) Block diagram of the feature extraction signal chain. (b) Schematic and illustration of the log-domain biasing generation. (c) Simplified schematic of the SAR ADC. }\label{ckt_gm_adc}
\end{figure}

Distinct brain oscillations exist in EEG with frequencies spaced logarithmically. Thus, energy extraction with frequency bins spaced in the logarithmic domain can achieve a high efficiency \cite{Liu2016}. In this work, each energy extraction channel consists of a 4th order stagger-tuned biquad filter, a Gilbert multiplier, and a leaky integrator, as shown in Fig. \ref{ckt_gm_adc} (a). The biquad filter can be programmed in 64 steps from 0.5Hz to 100Hz, generated logarithmically as illustrated in Fig. \ref{ckt_gm_adc} (b). The low and high cut-off frequencies can be independently programmed. A subthreshold biased Gm-C block was employed for an exponential I-V characteristic and high power efficiency. The variation in the subthreshold region is more prominent than the above threshold region. Instead of calibrating these mismatches, we corrected the induced errors by re-training the NN including the mismatches. This process was combined with the subject-specific training, which is required for achieving the optimal classification performance for each subject. The extracted features were digitized by the 10-bit SAR ADC. Fig. \ref{ckt_gm_adc} (c) shows a simplified schematic of the designed ADC \cite{CSR2016}.

Edge Machine learning algorithms have been deployed in neural implants for real-time classification \cite{JNE2021}. In this work, a novel two-stage NN architecture was proposed to enhance the performance of sleep stage classification and minimize the memory and computational costs. The first stage NN was trained using the features and labels from 1-sec segments, and the second stage NN was trained using the outputs from the first NN with labels from 30 sec data segments. Neither signals nor features need to be buffered for 30 sec, thus the memory requirement can be reduced. The first stage NN uses a feedforward architecture with three fully connected layers consisting of 32, 18, and 5 neurons, respectively. The second stage NN uses a long short-term memory (LSTM) layer. Unlike feedforward layers, a LSTM layer has recurrent connections with gating functions for capturing sequential information, which is important in sleep stage classification. All NN coefficients were quantized to 8 bits. Hardware friendly activation functions, $hard-sigmoid$ and $softsign$, were used as the gate and state activation functions, respectively.

\begin{figure}[!ht]
  \centering
  \includegraphics[width=0.5\textwidth]{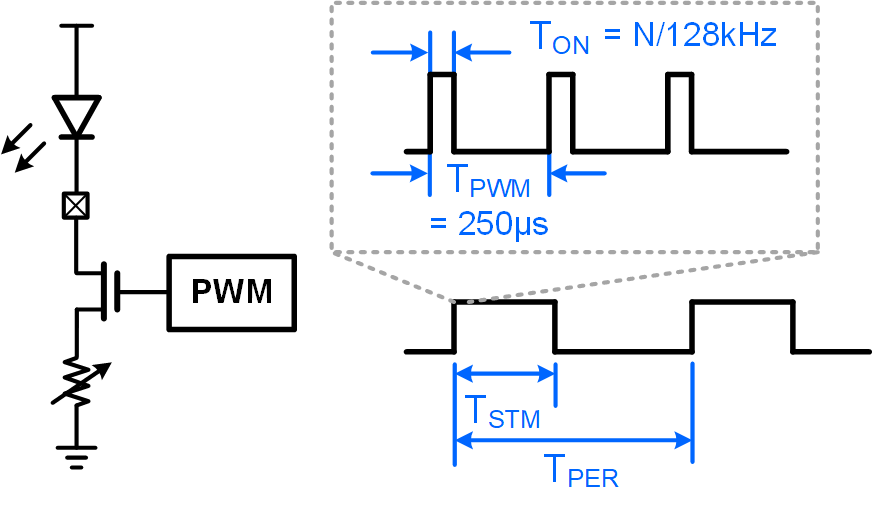}\\
  \caption{Schematic and timing diagram of the PWM based optical stimulator. }\label{ckt_stimulator}
\end{figure}

Four independent PWM based optical stimulators were designed in this work. The stimulators were connected to 16 I/O pads via a multiplexer for driving a 16-channel off-chip LED array. Fig. \ref{ckt_stimulator} illustrates the circuit design and stimulation timing diagram. The PWM refreshes at a constant rate of 4kHz, and the pulse width can be programmed in a step of 1/128kHz. On-chip registers and counters were designed to control stimulation with a programmable period TPER and a stimulation-ON time TSTM. The stimulator output stage consists of an NMOS pass device and a programmable resistor for limiting the peak current.

\section{Results}

The proposed mixed-signal circuit designs were simulated in Cadence. Monte-Carlo simulation was used to determine the variation. The input-output characteristics were modeled in Matlab. The circuit models and the two-stage NN based sleep stage classifier were trained by supervised learning in Matlab. The performance of the classifier was evaluated on a publicly available Sleep-EDF database (expanded) \cite{Kemp2000}. The sleep stages were manually annotated by experts.

\begin{figure}[!ht]
  \centering
  \includegraphics[width=0.6\textwidth]{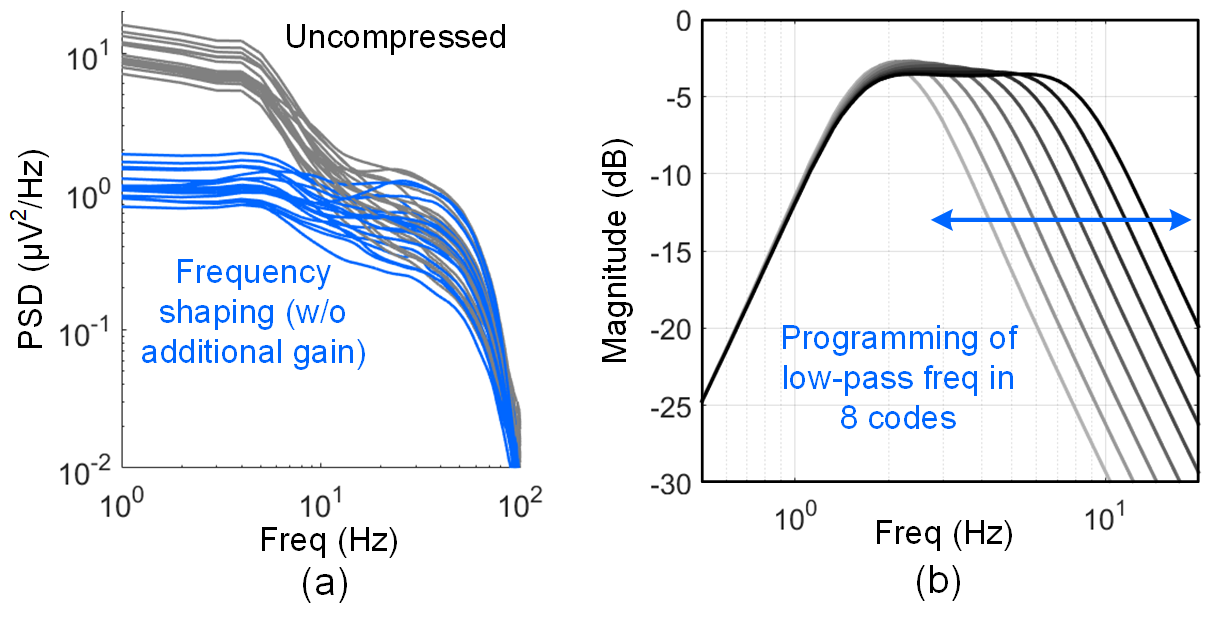}\\
  \caption{(a) Simulated PSD of 15 subjects before (grey) and after (blue) frequency shaping. No additional gain was added for illustration. (b) Simulated frequency responses of the biquad filter. }\label{exp_front-end}
\end{figure}

Table \ref{table_spec} summarizes the key specifications of the design.
The simulated input-referred noise of the EEG amplifier was 4.3$\mu$V in a bandwidth of 0.5-100Hz. Fig. \ref{exp_front-end} (a) shows the power spectral density (PSD) plots of the original and whitened EEG signals from 15 subjects before adding gain. The overall gain was designed to be programmable between 34-64dB (at 10Hz) in 16 steps, which was used for maximizing the SNR given the linear input range of the following energy extraction circuits. The designed SAR ADC achieves an ENOB of 8.81 in simulation.

\begin{table}[!ht]
\footnotesize
\centering
\caption{Summary of Key System Specifications}\label{table_spec}
\renewcommand{\arraystretch}{1.2}
\begin{tabular}{|c|c|}
\hline
Technology           & 180nm     \\ \hline
Supply Voltage       & 1.8V      \\ \hline
Recording ch \#      & 4         \\ \hline
Stimulator ch \#     & 16        \\ \hline
Neural Feat. Ex. \#  & 20        \\ \hline
1st stage NN \#      & 32/18/5   \\ \hline
2nd stage NN \#      & 30/5      \\ \hline
Input referred Noise & 4.3$\mu$V     \\ \hline
Bandwidth            & 0.5-100Hz \\ \hline
ADC ENoB             & 8.81      \\ \hline
SoC Power            & 97$\mu$W      \\ \hline
\end{tabular}
\end{table}

\begin{figure}[!ht]
  \centering
  \includegraphics[width=0.48\textwidth]{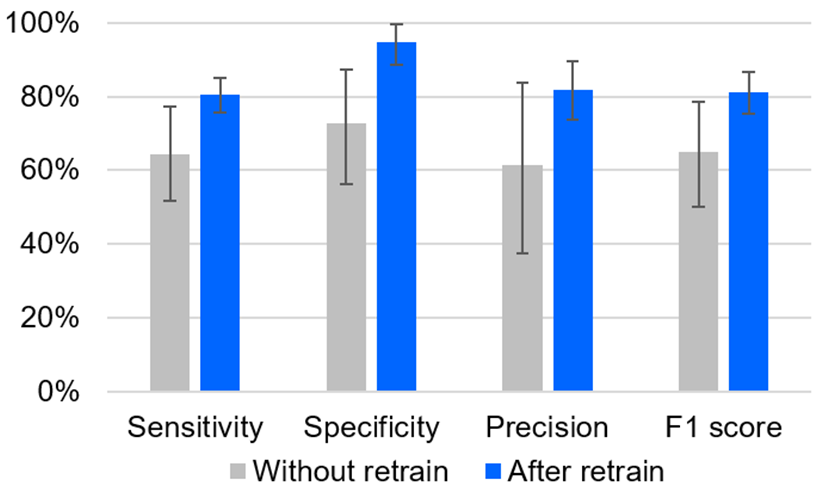}\\
  \caption{Improvement of re-training including circuits mismatches modeled from Monte-Carlo simulation.}\label{exp_mismatch}
\end{figure}

Fig. \ref{exp_mismatch} shows the performance improvement by NN retraining including mismatches modeled from circuits simulation. The results show that all performance merits were improved and variations were reduced after the retraining.

\begin{figure}[!ht]
  \centering
  \includegraphics[width=0.65\textwidth]{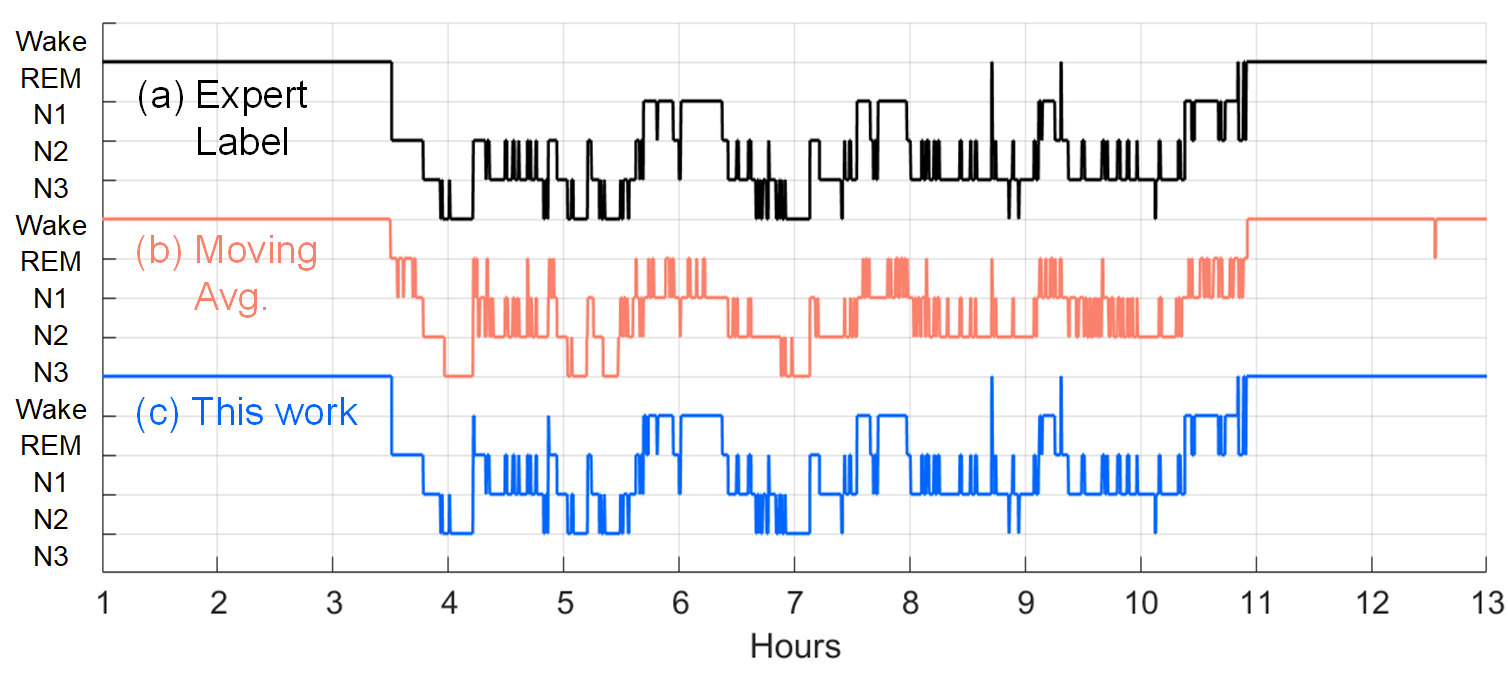}\\
  \caption{Sleep stage classification of one night. (a) expert annotation, (b) neural network with traditional moving average, and (c) the proposed 2 stage NN. }\label{exp_sleep_scoring}
\end{figure}

\begin{figure}[!ht]
  \centering
  \includegraphics[width=0.6\textwidth]{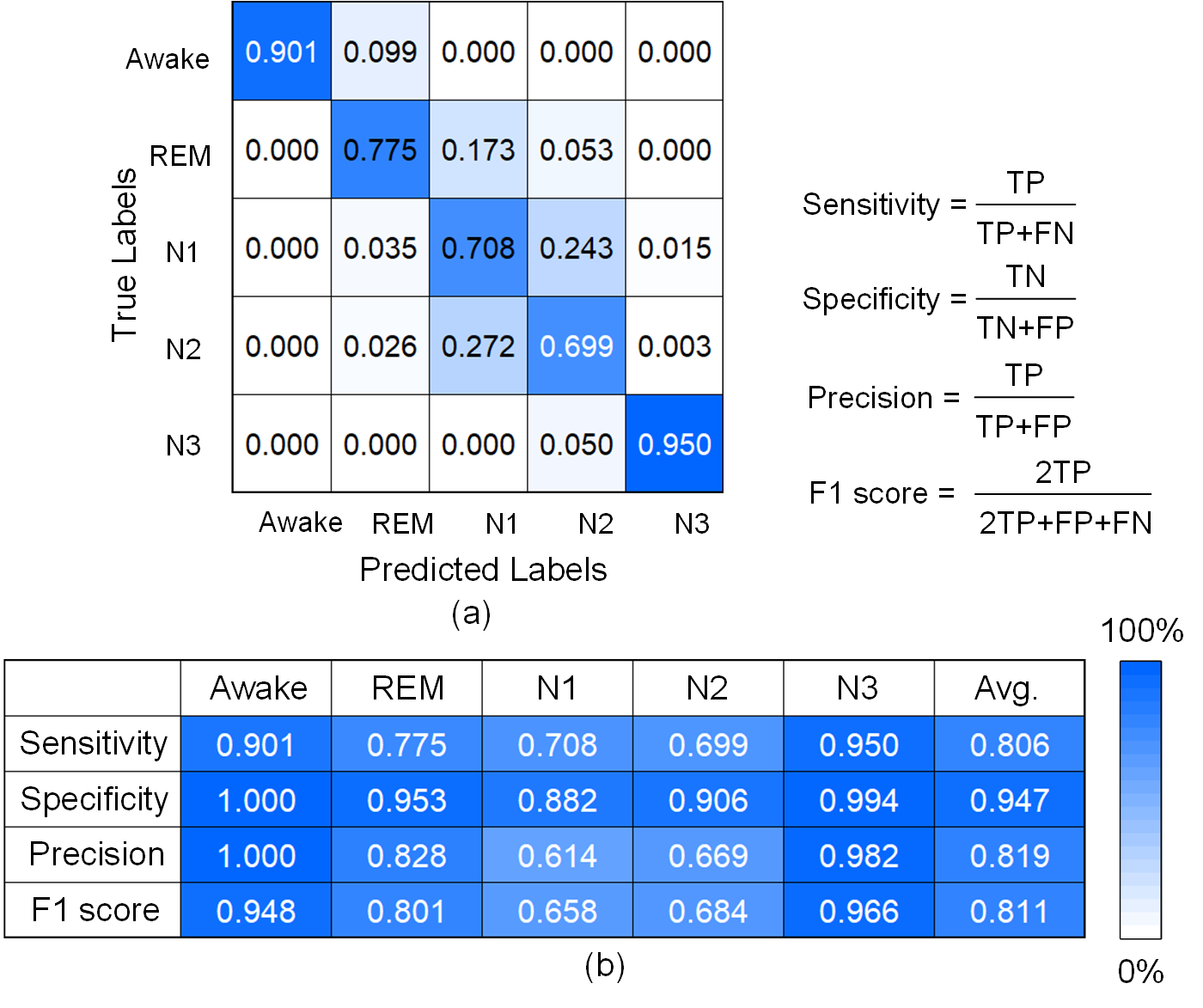}\\
  \caption{(a) Confusion table of the classification results, and (b) summary of the performance of 40 nights from 20 subjects.}\label{exp_accuracy}
\end{figure}

Fig. \ref{exp_sleep_scoring} plots one night of the sleep stage classification of expert annotation in comparison with traditional moving average and the proposed two-stage NN. To quantify the classification performance, standard metrics were used in 40 nights of recordings from 20 subjects. Fig. \ref{exp_accuracy} (a) and (b) shows the classification confusion table and performance metrics. In summary, the averaged sensitivity of the five sleep stages was 0.806, the averaged specificity was 0.947, averaged precision was 0.819, and the averaged F1 score was 0.811.

Table \ref{table_comparison} compares the performance with the current state-of-the-art. The proposed SoC design achieved high performance in real-time sleep scoring and closed-loop optogenetic intervention at an ultra low power consumption.

\begin{table*}[!ht]
\centering
\caption{Comparison with the State-of-the-art}\label{table_comparison}
\scriptsize
\renewcommand{\arraystretch}{1.2}
\begin{tabular}{|c|c|c|c|c|c|}
\hline
                                                                    & \begin{tabular}[c]{@{}c@{}}Imtiaz\\ 2017 \cite{Imtiaz2017}\end{tabular}                       & \begin{tabular}[c]{@{}c@{}}Kassiri\\ 2017 \cite{Kassiri2017}\end{tabular}                  & \begin{tabular}[c]{@{}c@{}}Liao\\ 2018 \cite{Liao2018}\end{tabular}                         & \begin{tabular}[c]{@{}c@{}}Phan\\ 2019 \cite{Phan2019}\end{tabular}                         & \begin{tabular}[c]{@{}c@{}}This \\ work\end{tabular}                        \\ \hline
\begin{tabular}[c]{@{}c@{}}Classifi-\\ cation\\ Stages\end{tabular} & \begin{tabular}[c]{@{}c@{}}5 stages\\ Awake/\\ REM/N1/\\ N2/N3\end{tabular} & \begin{tabular}[c]{@{}c@{}}3 stages\\ Awake/\\ REM/\\ NREM\end{tabular} & \begin{tabular}[c]{@{}c@{}}5 stages\\ Awake/\\ REM/N1\\ /N2/N3\end{tabular} & \begin{tabular}[c]{@{}c@{}}5 stages\\ Awake/\\ REM/N1/\\ N2/N3\end{tabular} & \begin{tabular}[c]{@{}c@{}}5 stages\\ Awake/\\ REM/N1/\\ N2/N3\end{tabular} \\ \hline
\begin{tabular}[c]{@{}c@{}}Signal \\ type\end{tabular}              & EEG                                                                         & EEG/EMG                                                                 & EEG                                                                         & EEG                                                                         & \begin{tabular}[c]{@{}c@{}}EEG/EOG/\\ EMG\end{tabular}                      \\ \hline
Features                                                            & FFT                                                                         & Filtering                                                               & Filtering                                                                   & Learned                                                                     & \begin{tabular}[c]{@{}c@{}}Energy\\ Features\end{tabular}                   \\ \hline
Algorithm                                                           & FSM                                                                         & Threshold                                                               & \begin{tabular}[c]{@{}c@{}}5-step \\ XGBoost\end{tabular}                   & CNN                                                                         & \begin{tabular}[c]{@{}c@{}}2-stage \\ NN\end{tabular}                       \\ \hline
Dataset  & 20 cases  & 9 mice  & 1 case  & 20 cases  & 20 cases  \\ \hline
Accuracy  & 98.7\%  & 81.7\%  & 86.3\%  & 82.3\%  & 80.6\%  \\ \hline
Sensitivity  & N/A  & 81.7\%  & N/A  & 74.3\%  & 80.6\%  \\ \hline
Specificity  & N/A  & 93.9\%  & N/A  & 95.1\%  & 94.7\%  \\ \hline
Precision   & N/A  & N/A  & 70.8\%$^*$  & N/A  & 81.9\%  \\ \hline
F1 score   & N/A  & N/A  & 66.4\%$^*$  & 74.7\%   & 81.1\%   \\ \hline
Hardware  & ASIC   & FPGA  & FPGA  & N/A  & ASIC  \\ \hline
Stimulation  & No   & Yes  & No  & No  & Yes  \\ \hline
Power  & 575$\mu$W  & 3.6mW  & 102$\mu$W  & N/A  & 97$\mu$W  \\ \hline
\end{tabular}
\end{table*}

\section{Conclusions}
The key contributions of this work include: 1) the first reported SoC design with a complete closed-loop pathway from PSG recording, sleep stage classification, to optical stimulation; 2) a novel two-stage NN architecture that enhances the classification performance while minimizing the computational and memory cost; 3) an ultra-low-power neural feature extraction design with mismatch correction from subject-specific off-line re-training; 4) an amplifier design with a pre-whitening technique for equalizing feature magnitude, maximizing dynamic range, and preventing amplifier saturation from motion artifacts. Future work will fabricate the SoC and test it in sleeping NHPs \cite{Richardson2017}.



\end{document}